# Localizing axial dense emitters based on single-helix point spread function and deep learning

Yihong Ji,[1,3] Danni Chen,[1,3,*] Hanzhe Wu,[1] Gan Xiang,[1] Heng Li,[2] Bin Yu,[1] Junle Qu[1]

[1]*College of Physics and Optoelectronic Engineering, Key Laboratory of Optoelectronic Devices and Systems of Ministry of Education and Guangdong Province, Shenzhen University, Shenzhen 518060, China*

[2]*Tsinghua-Berkeley Shenzhen Institute (TBSI), Tsinghua University, Shenzhen, 518055, China*

[3] *contributed equally to this work*

\* *danny@szu.edu.cn*

**Abstract:** Stimulated Emission Depletion Microscopy (STED) can achieve a spatial resolution as high as several nanometers. As a point scanning imaging method, it requires 3D scanning to complete the imaging of 3D samples. The time-consuming 3D scanning can be compressed into a 2D one in the non-diffracting Bessel-Bessel STED (BB-STED) where samples are effectively excited by an optical needle. However, the image is just the 2D projection, i.e., there is no real axial resolution. Therefore, we propose a method to encode axial information to axially dense emitters by using a detection optical path with single-helix point spread function (SH-PSF), and then predicted the depths of the emitters by means of deep learning. Simulation demonstrated that, for a density 1~ 20 emitters in a depth range of 4 μm, an axial precision of ~35 nm can be achieved. Our method also works for experimental data, and an axial precision of ~63 nm can be achieved.

## 1. Introduction

STED is a super-resolution optical microscopy technique. By superimposing a doughnut depletion beam on a Gaussian excitation beam, the excited fluorescent molecules in the non-central region can be stimulated to return to the ground state by stimulated emission instead of fluorescence emission, with a probability proportional to the local intensity of the depletion beam. Then, a narrowed PSF of fluorescence image will be generated. Because a depletion beam with higher intensity means a narrower PSF, the resolution can be improved to even several nanometers [1]. In order to maintain the resolution at deep depth, Gaussian-Bessel STED (GB-STED) microscopy was proposed, where a hollow Bessel beam with the distortion-free nature was used as the depletion beam, so a doughnut structure required could be maintained in an extended DOF [2]. In order to induce severe out-of-focus noise, the excitation strategy adopted in GB-STED is a conventional Gaussian beam which has a axially confined profile to avoid unnecessary excitation outside the axial detection region. Accordingly, the confocal detection mode has to be used, so a 3D scanning is required for a 3D volumetric image. Such a 3D scanning sacrifices the temporal resolution and does not make full use of the non-diffraction properties of the depletion beam. A more efficient strategy is to apply non-diffraction modes on both excitation and depletion beams, which can be named as BB-STED [3]. With proper parameters, emitters in a laterally confined and axially extended volume can be excited effectively, which means an effective optical needle can be generated. Actually, such an optical needle has been applied in light sheet microscopy to increase both the field of view and the axial resolution [4]. However, when the strategy is used in a traditional microscopy, there is a main obstacle that it is hard to retrieve axial information of dense emitters excited in the optical needle, because images of all excited emitters, shown as spots, overlapped severely with each other. Regarding this issue, there are several solutions that can be referred, including using spiral PSF such as double-helix PSF (DH-PSF) [5] or single-helix PSF (SH-PSF) [6]. The spiral PSF methods spread spots of emitters at different depths to different azimuths. DH-PSF has been used in single molecule localization microscopy (SMLM) [5] and particle tracking [7]. However, if the emitters in the optical needle are too dense axially, the dense spots are still too close to resolve. This problem is similar with dense emitter localization in SMLM, so algorithms used in SMLM might be referable solutions, such as Compressed Sensing (CS) [8], Dominion Astrophysical Observatory Stochastic Optical Reconstruction Microscopy (DAOSTORM) [9]. Besides these algorithms, there is an emerging solution, i.e., deep learning.

Deep learning is a major branch of machine learning, which is an algorithm that performs multi-level analysis and computation on data based on sufficient samples by constructing multiple neurons with a large structure. Deep convolutional neural networks (CNNs) have been widely applied in different research fields, including image classification [10], segmentation [11], and other areas. In image recognition, deep learning automatically extracts features from input images by constructing complex and deep network models. The continuous development of deep learning has made it widely used in various fields, and it also has many applications in super-resolution microscopy imaging. For example, Deep-STORM [12] shows significantly higher speed than existing approaches, which is the first to apply deep learning to single-molecule localization, and ANNA-PALM [13], which can generate super-resolution STORM images directly from a wide-field image using a generative adversarial network [14]. Obviously, deep learning is also widely used in the field of 3D super-resolution microscopy imaging, such as DeepSTORM3D [15], which can achieve large axial range modification PSF localization. In order to reduce the number of original images required for reconstruction, the U-Net-SIM3 network reduces the number of original images required for 3D SIM reconstruction to 1/5 of the original [16]. The powerful learning ability of deep learning has greatly improved the spatial and temporal resolution of super-resolution imaging.

Here, based on SH-PSF and deep learning, we propose a method (named as SH-DL) of simultaneously resolving dense emitters excited in an optical needle. The emitters are imaged using a detection system with SH-PSF. The Encoder-Decoder network architecture in deep learning algorithm is used to resolve the overlapped spot-like images of these emitters. Simulation demonstrated that, for a density up to 20 emitters in an effective range of depth, axial positioning precision of ~35 nm can be achieved. Experiment results from a designed sample shows that for an axial emitter density of 20, the axial localization precision is ~ 63 nm. Since all axial information of excited emitters in an optical needle can be retrieved with this method, we believe that combined with a 2D scanning of the optical needle excitation, the final volumetric image with 3D nano-resolution will be easily reconstructed.

## 2. Method

The principle of the SH-DL method can be depicted with Fig.1. As is shown in Fig.1a, under the excitation of an optical needle, dense emitters in the axial direction are excited, and then imaged using a detection optical path with SH-PSF. Each raw image consists of overlapping spots which are images of the emitters. The images are then analyzed using deep learning shown in Fig.1b.

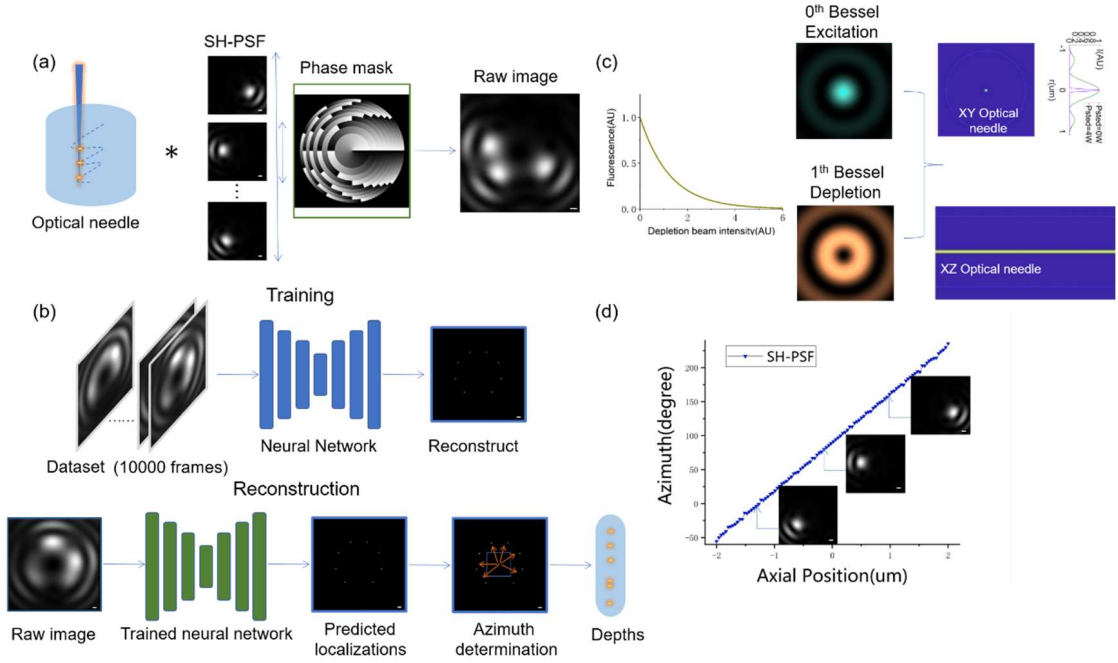

Fig.1. The principle of the SH-DL method. (a) Raw image generation in the SH-DL method. When excited with an optical needle, the excited emitter distribution is the product of the intensity of the optical needle and the emitter distribution. The raw image is the convolution of the excited emitters distribution and the SH-PSF. (b) Analytical localization of dense spots using neural network learning, including model training (upper) and raw image processing with trained neural network (lower) (c) Generation of nano-optical-needle by means of the BB-STED. (d) Calibration curve of spot azimuth versus emitter depth, and three images of a single emitter at three depths are inserted. Scale Bar:200nm.

Optical needles with diffraction-free characteristics and super-resolution width can be achieved using the BB-STED [3]. The lateral spatial resolution of our method is dependent on the size of the optical needle. AS is shown in Fig.1c, for emitters with the characteristic curve of fluorescence decreasing with the depletion beam intensity, which are excited with a 0th Bessel beam and depleted with a concentric 1st Bessel beam, an optical needle can be generated and its intensity profiles in xy plane and xz plane demonstrate that the optical needle is tens of nanometers in width and several microns in length.

The SH-DL method uses a detection optical path with SH-PSF instead of DH-PSF here because of two reasons. Actually, similar with DH-PSF, SH-PSF encode depths of emitters with spot azimuths (Fig.1d), but the latter has only one lobe, so the raw image with SH-PSF has higher signal-noise-ratio (SNR). One-lobe brings one more advantage that, the effective depth range of the SH-PSF is twice of that of the DH-PSF, because the depths can be encoded in an azimuth range of $2\pi$ in SH-PSF, while in DH-PSF consisting of two lobes, it is only $\pi$. SH-PSF can be implemented in various ways, including superposition of certain Laguerre-Gaussian modes [5], and the Fresnel zone plate method [17] which was used for the following numerical simulations. The Fresnel zone plate can adjust the effective axial positioning range by adjusting the number of phase rings, so the number of emitters per micron in z axis is not proper for indicating the axial resolving ability here. In the following discussion, we use the number of spots at different azimuths, $n$, within an azimuthal range of $5\pi/3$, as the indicator of axial emitter density. To avoid confusion of the spots at azimuth $2\pi$ and 0, the azimuthal range is slightly smaller than $2\pi$. The corresponding depth range is defined as the effective axial range. As is shown in Fig.1a, the 2D image consisting of single-helix spots is the convolution of emitters excited by the optical needle and the SH-PSF.

The SH-DL uses deep learning to locate the spots. In the model training stage (upper part in Fig. 1b), a simulated dataset is input to the neural network, to learn localizing dense single-helix spots. Then, as is shown in the lower part of Fig.1b, the raw image is put into the trained and optimized neural network to predict the localization of the spots. The azimuths of these localizations relative to the center of the optical needle are calculated. Finally, the 3D reconstruction is performed by translating predicted localizations into 3D Cartesian coordinates $(x_{needle}, y_{needle}, z_n)$, where $(x_{needle}, y_{needle})$ is the coordinates of the center of the optical needle, and $z_n$ represents the depths decoded according to the calibration curve shown in Fig.1d.

### 3. Simulation

The excitation light path scheme is the BB-STED [3], where the wavelength of the 0th-order Bessel beam for excitation is set to 635 nm, and that of the 1st-order Bessel beam for depletion is set to 750 nm. The maximum intensity of the depletion beam is set to 10 times of that of the excitation beam. As is shown in Fig.1.c, based on the fluorescence characteristic of organic dye Atto647N, the effective fluorescence intensity varies with the depletion beam power, so an effective optical needle with transverse full width at half maximum (FWHM) of ~50 nm and a length of 10 μm can be generated. There is a pair of side lobes at ~ ±1 μm, but their intensity is only 6% of that at the center, so only the center lobe is considered in following simulations.

The detection optical path with SH-PSF used for numerical simulation is similar with that used in [18]. The effective axial range is set to 4 μm. The effective pixel size of the final raw images is set to 10 nm.

*3.1 Dataset*

In the process of generating dataset, in order to take advantage of deep learning in image reconstruction, we need to generate images of samples consisting of extremely dense emitters in z axis. Therefore, the maximum axial density set here is up to 20. 1~20 emitters were randomly distributed within a cylindrical volume whose diameter was set to 100 nm and axial height was set to the effective axial range. Samples were excited with the optical needle and imaged by the optical system with SH-PSF. Gaussian noise with 0.1 mean and 0.01 variance, and Poisson noise were added to all raw images after normalization. Finally, 10,000 frames of raw images with a size of $208 \times 208$ were generated, of which 7,000 frames and 3,000 frames were used as the training and the validation set respectively.

*3.2 Network architecture*

In terms of deep learning algorithms, inspired by the previous works of Deep-STORM [12], we adopted the Encoder-Decoder architecture based on CNNs. The Encoder-Decoder architecture is popularly used in the field of machine translation, and it is also well-suited for image segmentation and reconstruction. The network is mainly divided into two modules: the down-sampling module, and the up-sampling module. The former extracts higher-level features through layer-by-layer networks, while the latter maps the features to higher-resolution images. The high-level feature map outputs higher-resolution feature maps through layer-by-layer networks, and finally reconstructs the localization image.

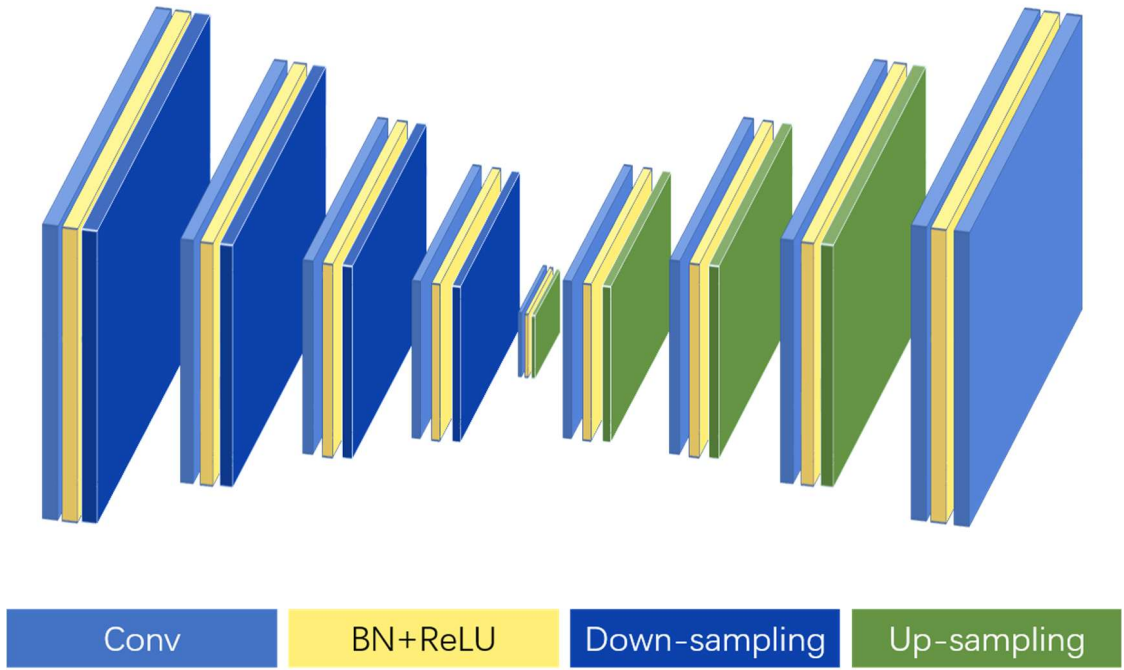

Fig.2. The architecture of the neural network

The network structure is shown in the Fig.2. The size of the square in the figure represents the input size of the image. The initial image input size is $208 \times 208$. The network consists of four main parts.

1. Conv: The convolution kernel size of the convolutional layer is $3 \times 3$ to generate feature maps. Initially, we set 64 filters to extract features from the image. In the down-sampling module, the number of filters is doubled for each subsequent convolutional layer. In the up-sampling module, the number of filters is halved for each subsequent convolutional layer. The size of the convolved image remains unchanged, and corresponding feature maps are generated after passing through the convolutional layers.

2. BN + ReLU: Batch normalization (BN) is usually applied after the convolution layer to normalize the feature maps after convolution, and to satisfy the distribution rule of mean 0 and variance 1 [19]. This accelerates the training process of the entire network. Then, the rectified linear units (ReLU) [20] activation function layer provides nonlinearity, and the ReLU function will turn all negative values passed in to 0, further reducing the sparsity of the network and preventing overfitting.

3. Down-sampling: We use Max Pooling for down-sampling, with a pooling matrix of $2 \times 2$. It takes the maximum element of the pixels in the $2 \times 2$ region as the corresponding pixel value of the pooled image, so the size of the pooled image is reduced by half each time to discard redundant information. Max Pooling is beneficial for maintaining the invariance of image features and further controlling overfitting.

4. Up-sampling: We use transposed convolution with training parameters to perform up-sampling, which is more conducive to generating higher-quality high-resolution images. The convolution kernel size of transposed convolution is $3 \times 3$. By setting strides=2, the image size after each transposed convolution during the up-sampling process is doubled, gradually restoring to the size we need.

Finally, a convolutional layer with a linear activation function is used for prediction. The network contains a total of approximately 25M training parameters.

*3.3 Loss function*

We optimized our model prediction results by setting a loss function. During forward propagation, parameters are passed through the CNN layer by layer, and finally passed into the loss function to calculate the loss. In backward propagation, parameters are returned from the loss function to each layer of the network, and the loss function is optimized through an iterative algorithm. Therefore, we need to set our loss function according to the desired goal. Our loss function consists of a data-fidelity term and one regularization term, the first part is the mean squared error loss, and the second part is the L1 regularization. The loss function expression is as follows:

$$Loss = \frac{1}{N}\sum_{i=1}^{N}\|\tilde{y}_i * g - y_i * g\|_2^2 + \lambda\|\tilde{y}_i\|_1 \tag{1}$$

Assuming that the number of images in the training set is N, $\tilde{y}_i$ is the predicted value of our model, and $y_i$ is the ground-truth (GT). We used the mean squared error loss to measure the gap between the true value and the predicted value, and they are convolved with a $5 \times 5$ size Gaussian Kernel g. At the same time, we used the l1 norm for regularization, where $\lambda$ is the regularization parameter, which is usually set to 1. By using the l1 norm, we can create a certain sparsity in the solution space, which is beneficial for automatically selecting the features we need [21].

We used the Python language and built networks based on the Keras [22] with a TensorFlow [23] backend framework. Due to the need to achieve higher density localization, we set 64 samples as the training data for each batch and used the Adam [24] optimization algorithm for feedback. We implemented the training on a high-performance GPU workstation, using a NVIDIA GeForce RTX3090Ti with 24G of video memory. The entire training process took approximately 2 hours.

*3.4 Model validation*

We simulated images as a test dataset for model performance evaluation. As is shown in Fig.3, the GT and predicted localizations of the spots are highlighted with green circles and red crosses respectively, and projected on the corresponding images with different spot density ($n = 1 \sim 20$). As is shown in Fig.3a~j, for low spot density, the predicted localizations coincide with the GT ones quite well. For example, in Fig.3c ($n$=3), the predicted localizations of the three spots are slightly offset from the GT localizations by 10 nm, 20 nm, and 20 nm. The corresponding errors in the final axial positions of the three emitters are 21 nm, 18 nm, and 7 nm, respectively. As the density increases (Fig.3k~t), the deviation increases, indicating a decrease in localization accuracy and consequently a decreased accuracy in the next axial position prediction.

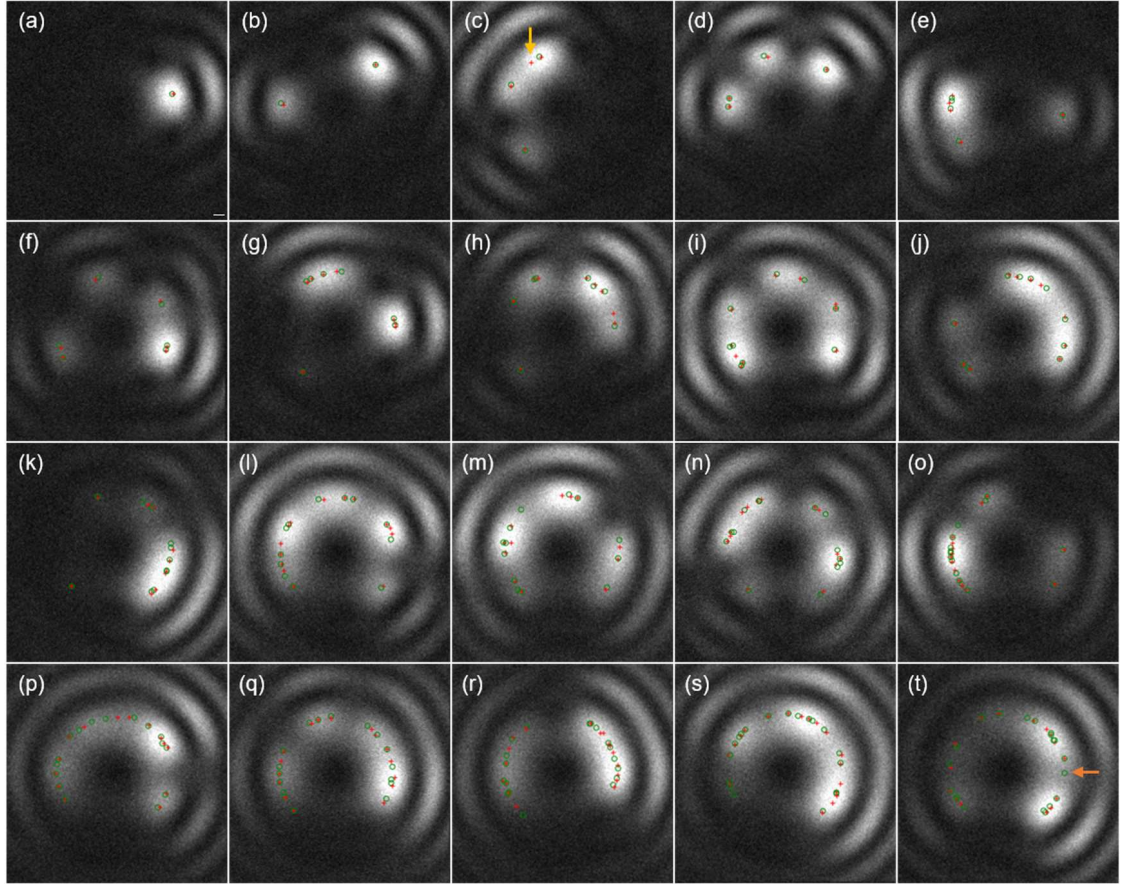

Fig.3.  Spot localizing for raw images with different spot density. (a-t) Raw images with spot density from 1 to 20. Green circles represent corresponding GT localizations of the spots, while the red crosses represent predicted ones. Scale bar: 50 nm.

Next, we statistically analyzed the predicted localizations to estimate the localization accuracy and precision under different spot density $n = 1 \sim 20$. For each spot density $n$, 500 images with $n$ spots but random azimuth distribution were generated. Then all predicted localization in the 500 images were paired with their closest GT localizations, and their relative offsets from their paired GT localization were calculated. The distributions of the offsets in x and y axis are shown in Fig.4a and Fig.4b respectively. Histograms were fitted with Gaussian curves, and then $< x_{offset} > = m_x \pm s_x$ nm and $< y_{offset} > = m_y \pm s_y$ nm were determined. The two mean values ($m_x$ and $m_y$) and two standard deviations (STDs) ($s_x$ and $s_y$) represent the spot localization accuracy and precision respectively. They are plotted in Fig.4c~f and then fitted with line equations. As the axial emitter density increases from 1 to 20, the accuracy of localization decreases a little from 21 nm to 28 nm (x axis, Fig.4c) and 21 nm to 24 nm (y axis, Fig.4d), while the precision decreases from 21 nm to 28 nm (x axis, Fig.4e) and 22 nm to 24 nm (y axis, Fig.4f).

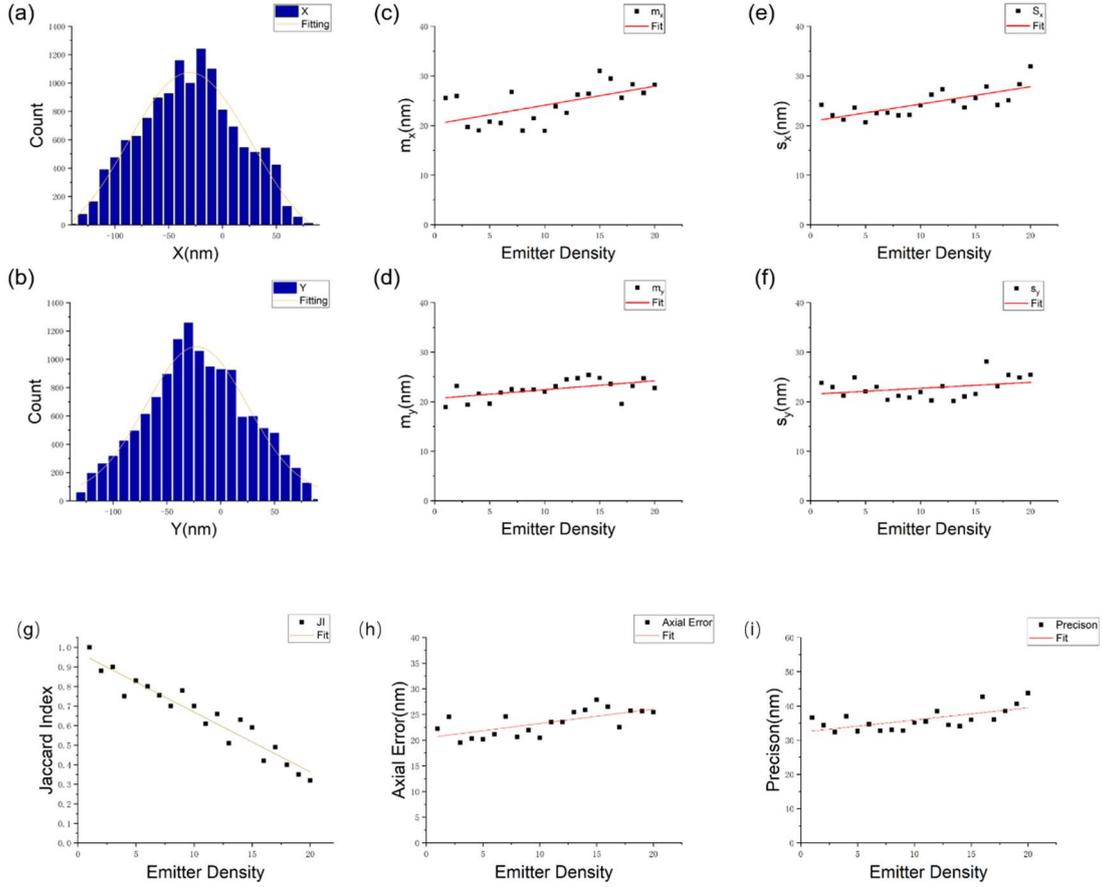

Fig.4. Statistical analysis of spot localization. (a, b) Histograms of the distributions of the offsets in x and y axis (for $n = 15$). The mean and STD values are then calculated form their fitted curves. (c~f) Plots of mean value (c,d) and STD (e, f) versus spot density. (g) Plot of Jaccard index versus emitter density. (h) Plot of axial error versus emitter density. (i) Plot of axial precision versus emitter density. All data in c~i are fitted with line equations.

Considered that the reconstructed axial structure is composed by predicted localizations, the predicted depths of the emitters were analyzed statistically and shown in Fig.4h and Fig.4i. As the emitter density increases from 1 to 20, the axial error increases from 22 nm to 25 nm, while the axial precision decrease from 36 nm to 43 nm.

It should be noted that unexpectedly situations may occur occasionally, including unexpected localizations (such as the localization pointed by a downwards arrow in Fig.3c) and localization of some spots being lost (such as the GT localization pointed by a leftwards arrow in Fig.3t). So, in order to evaluate the error of deep learning for single helix spot localization, the Jaccard Index (JI) [25] was introduced to analyze the predicted localizations used in Fig.4. The main calculation formula of JI is,

$$JI = \frac{TP}{FN+FP+TP} \qquad (2)$$

Where True positives (TP) represent correct localizations (lateral offset < 20 nm), false positives (FP) represent fault localizations (lateral offset > 20 nm), and false negatives (FN) include spots that are failed to be localized and localizations out of none spot. As is shown in Fig.4g, the Jaccard Index is plotted with respect to axial emitter density, and then fitted with a line equation. As the axial emitter density increases, the difficulty of prediction increases, and the Jaccard Index decreases to some extent. However, for emitter density as high as $n = 15$, JI is still greater than 0.5, which indicate that the localization results are still reliable even for high spot density.

*3.5 3D simulation reconstruction*

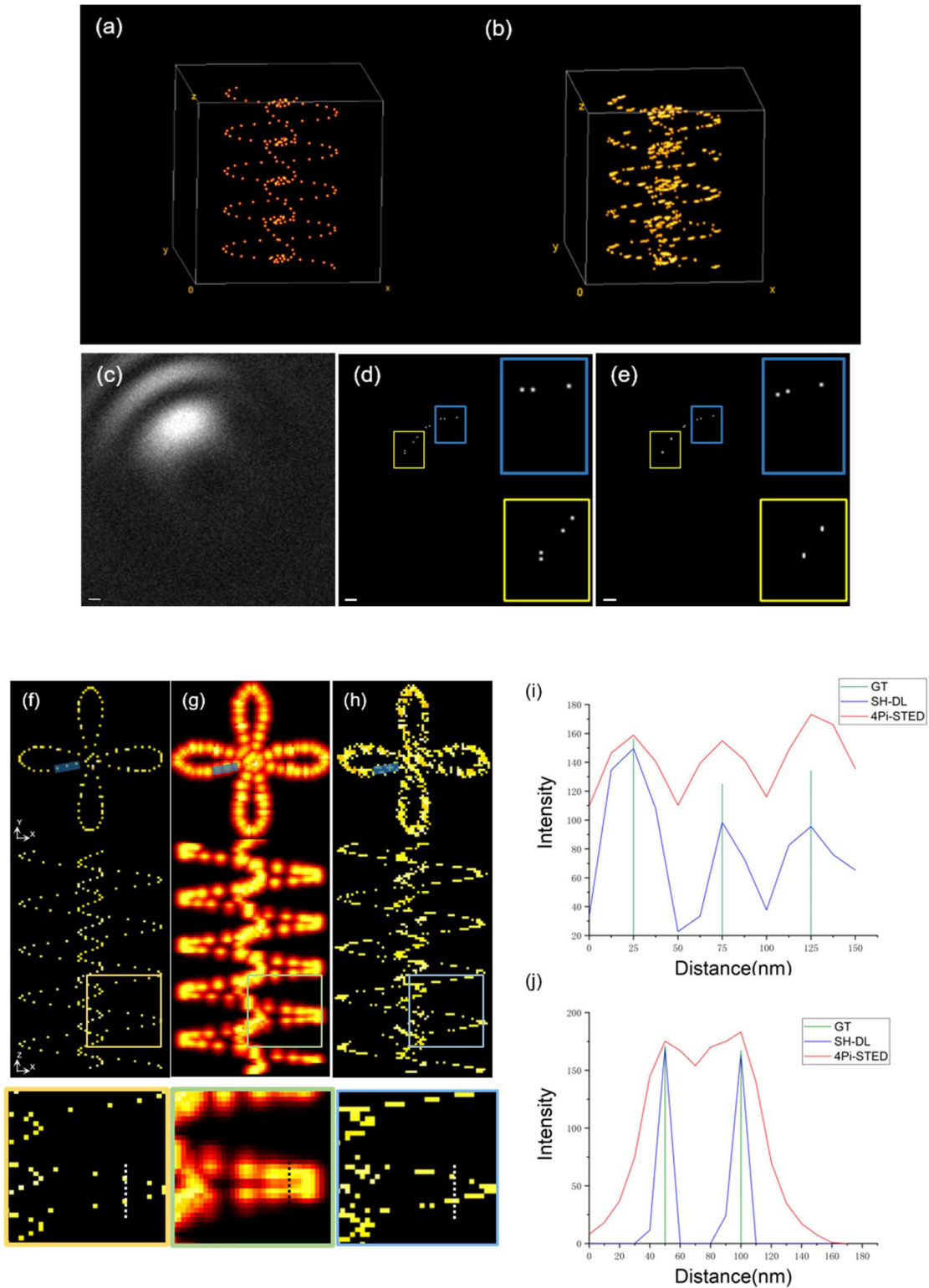

Fig.5. Reconstructed image of a 3D sample. (a~b) Emitter distribution (a) and reconstructed 3D structure (b) of the sample. (c~e)

One raw image (c) and its corresponding GT and predicted localizations (d and e respectively), where the inserted images are enlarged view of the square areas highlighted with corresponding colors. Scale Bar: 50 nm. (f~h) Projected GT emitter distribution (f), the 4Pi-STED image (g), and the SH-DL image (h) on the xy and the xz coordinate plane. Enlarged view of the square area from (f~h) are shown below correspondingly. (i~j) The intensity line profile along the approximately horizontal lines and vertical dashed lines in (f~h).

Next, a three-dimensional sample with $50 \times 50 \times 100$ pixels and voxel size of $10 \times 10 \times 10$ nm$^3$ was simulated (Fig.5a) and reconstructed (Fig.5b). First, a series of raw images of the sample were obtained by 2D scanning the sample with an optical needle of 50 nm (FWHM), with a step size of 20 nm. Fig.5c shows one of the raw images. Its corresponding GT and predicted localizations are shown in Fig.5d and Fig.5e respectively. The reconstructed structure shown in Fig.5b was projected on the xy and the xz coordinate plane and hereinafter referred to as the SH-DL image (Fig.5h). For comparison, a 4Pi-STED image of the same sample was simulated (Fig.5g) by convoluting the emitter distribution with an isotropy PSF simplified as a 3D Gaussian function with FWHM of 50 nm. Both the SH-DL and the 4Pi-STED images show reliable structure of the sample. However, the SH-DL image presents higher resolution. Fig.5i shows the intensity line profile along the approximately horizontal lines in Fig.5f~h. Compared with the GT structure, both the SH-DL and the 4Pi-STED image show comparable broadening of ~50 nm. The results are reasonable since the two methods have the same lateral resolution fundamentally. As to the intensity line profile along the vertical dashed lines, the 4Pi-STED can hardly resolve two layers separated by 50 nm in z axis, while two peaks can easily be identified by the SH-DL (Fig.5j). It demonstrates that the SH-DL image has better axial resolution than the 4Pi-STED image. It should be noted that, the SH-DL is also better than the 4Pi-STED in temporal resolution, because generally, the 3D scanning in the 4Pi-STED will take more time than the 2D scanning in the SH-DL.

## 4. Experimental validation

Designed experimental samples with known emitter distribution (See supplementary materials for details) were used to verify our method. Images of designed samples with random distribution were used to train the neural network which was then used to deal with the experimental samples. Raw images of four designed samples consisting of 9, 15, 10, and 19 emitters when excited with an optical needle (FWHM 50 nm) were generated and shown in Fig.6a~d respectively, where the GT and the predicted localization are marked with green circles and red crosses respectively. The final reconstructed results are shown in Fig.6e~h respectively. Furthermore, the Jaccard Index was also used to evaluate the performance of our method in analyzing real experimental results of different axial emitter densities (Fig. 6i). Compared with the simulated results, the coefficient from the experimental results shows a similar downward trend with increasing emitter density, but is generally a little worse. Nevertheless, it remains at a good level of ~ 0.5 at $n = 10$. The final axial positioning results were also calculated and evaluated. As is shown in Fig.6j and Fig.6k, compared with the curves of simulations, both the accuracy and precision of the experimental results experience a similar but rapider downward trend with increasing emitter density. The fitted lines in Fig.6j and Fig.6k shows that, as the axial emitter density increases from 1 to 20, the axial localization error increases from 21 nm to 48 nm, and the precision decreases from 34 nm to 63 nm.

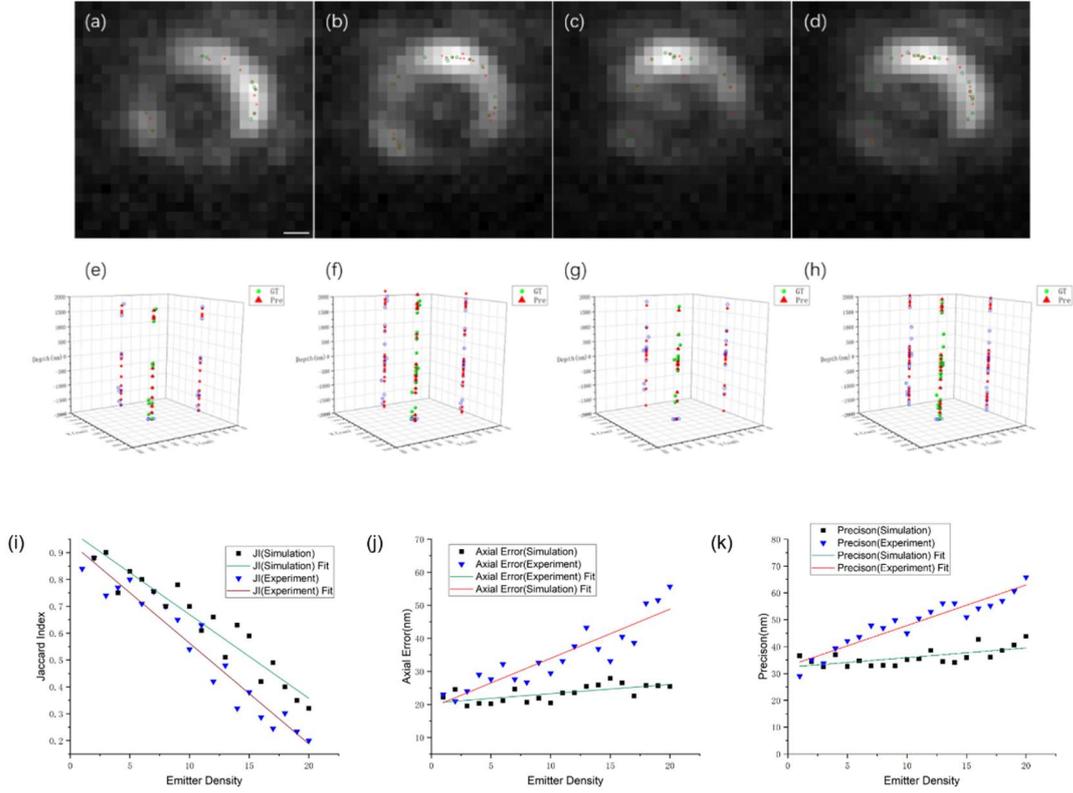

Fig.6. Experimental validation of the SH-DL with designed samples. (a-d) Raw images of four samples consisting of 9, 15, 10, and 19 beads at different depths, respectively. GT and predicted localizations are marked with green circles and red crosses respectively. scale bar: 208 nm. (e-h) Three-dimensional projections of the GT beads and predicated localizations in (a-d). GT and predicted localizations are marked with green spots and red triangles respectively. (i ~ k) Plots of Jaccard Index (i), axial error (j) and axial precision (k) versus emitter density based on experimental data and simulations. All data in i~k are fitted with line equations.

## 5. Conclusion

Here, based on SH-PSF and deep learning, we propose the SH-DL method to locate multiple emitters which are densely distributed along z axis. First, these emitters are imaged with a detection system whose PSF is engineered to SH-PSF, so the emitters at different depths are simultaneously imaged as spots at different azimuths. For axially dense emitters, the spots are overlapped densely. A neural network is trained and then used to located the azimuths of the spots. Finally, these localizations are translated into axial positions of the emitters, according to the spot azimuths. The SH-DL is useful when the excited emitters are confined laterally, and such optical needle can be realized when the sample is excited by means of the BB-STED. Simulations demonstrated that combined with an excitation optical needle of 50 nm, JI is still greater than 0.5 for emitter density as high as $n = 15$. After a 2D scanning, the 3D structure of a simulated sample was reconstructed with lateral resolution as high as 50 nm, and emitters separated by 50 nm can easily resolved by the SH-CS. Experimental data validated the feasibility of the SH-DL as well. For emitter density as high as 20, the axial localization precision is 63 nm. It should be noted that FWHM of the optical needle for simulations and experimental validation is 50 nm, and a smaller optical needle will improve the resolution in three dimensions solidly. We believe that the SH-DL will become a useful tool for rapid volumetric imaging with 3D nano-resolution in the future.

## 6. Acknowledgments

This work was supported by National Key Research and Development Program of China (2022YFF0712500), National Natural Science Foundation of China (Grant Nos. 11774242, 62175166, 61335001), Shenzhen Science and Technology Planning Project (Grant No. JCYJ20210324094200001, JCYJ20200109105411133).

## 7. References

1. Hell SW, Wichmann J.,"Breaking the diffraction resolution limit by stimulated emission: stimulated-emission-depletion fluorescence microscopy." Optics Letters 19(11). ; ,780-782(1994).
2. W. Yu, Z. Ji, D. Dong, X. Yang, Y. Xiao, Q. Gong, P. Xi, K. Shi., "Super-resolution deep imaging with hollow Bessel beam STED microscopy." Laser Photonics Rev10(1), 147-152(2016).
3. P. Zhang, P. M. Goodwin, and J. H. Werner, "Fast, super resolution imaging via Bessel-beam stimulated emission depletion microscopy," Opt. Express 22(10), 12398-12409 (2014)
4. Lu, C.-H. et al., "Lightsheet localization microscopy enables fast, large-scale, and three-dimensional super-resolution imaging." Commun. Biol 2, 1–10(2019).
5. Pavani, S. R. P.; Thompson, M. A.; Biteen, J. S.; Lord, S. J.; Liu, N.; Twieg, R. J.; Piestun, R.; Moerner, W. E., "Three-Dimensional, Single-Molecule Fluorescence Imaging Beyond the Diffraction Limit By Using a Double-Helix Point Spread Function. Proc." Natl. Acad. Sci. U. S. A106, 2995– 2999(2009).
6. Matthew D. Lew, Steven F. Lee, Majid Badieirostami, and W. E. Moerner, "Corkscrew point spread function for far-field three-dimensional nanoscale localization of pointlike objects," Opt. Lett36, 202-204 (2011)
7. Chen, Danni, Li, Heng, Yu, Bin and Qu, Junle. "Four-dimensional multi-particle tracking in living cells based on lifetime imaging." Nanophotonics, 11,(8),1537-1547(2022).
8. ZHANG S W, CHEN D N, NIU H B. "3D localization of high particle density images using sparse recovery.". Applied Optics 54(26),7859-7864(2015).
9. HOLDEN, S. J., UPHOFF, S. & KAPANIDIS, A. N. "DAOSTORM: an algorithm for high-density super-resolution microscopy." Nature Methods 8(4), 279-280(2011).
10. A. Krizhevsky, I. Sutskever, and G. E. Hinton, "ImageNet classification with deep convolutional neural networks," in Proceedings of the 25th International Conference on Neural Information Processing Systems, F. Pereira, C. J. C. Burges, L. Bottou, and K. Q. Weinberger, eds. (Curran Associates Inc.1097–1105(2012).
11. O. Ronneberger, P. Fischer, and T. Brox, "U-net: convolutional networks for biomedical image segmentation," in International Conference on Medical Image Computing and Computer-Assisted Intervention, N. Navab, J. Hornegger, W. M. Wells, and A. F. Frangi, eds. (Springer), pp. 234–241(2015).
12. E. Nehme, L. E. Weiss, T. Michaeli, and Y. Shechtman. "Deep-STORM: Super Resolution Single Molecule Microscopy by Deep Learning."Optica 5(4), 458-464(2018).
13. Ouyang W, Aristov A, Lelek M, Hao X, Zimmer C.,"Deep learning massively accelerates super-resolution localization microscopy." Nat Biotechnol 36(5), 460–8(2018).
14. Ian Goodfellow, Jean Pouget-Abadie, Mehdi Mirza, Bing Xu, David Warde-Farley, Sherjil Ozair, Aaron Courville, and Yoshua Bengio. "Generative adversarial networks." Commun. ACM 63, 11, 139–144(2020).
15. E. Nehme et al., "DeepSTORM3D: dense 3D localization microscopy and PSF design by deep learning." Nat. Methods 17, 734 (2020)
16. Jin L, Liu B, Zhao F, Hahn S, Dong B, Song R, et al. "Deep learning enables structured illumination microscopy with low light levels and enhanced speed." Nat Commun 11(1), 1934(2020).

17. Zhao Zhong-Chao, Yang Xu-Feng, et al. "Point spread function of incoherent digital holography based on spiral phase modulation." Acta Physica Sinica 67(1), 014203-014203(2018).
18. Li HF, Wang FM, et al. "Particles 3D tracking with large axial depth by using the 2π-DH-PSF." Optics Letters(46), 5088-5091(2021).
19. S. Ioffe and C. Szegedy, "Batch normalization: accelerating deep network training by reducing internal covariate shift," in Proceedings of the 32nd International Conference on Machine Learning, F. Bach and D. Blei, eds. (JMLR.), 448–456(2015).
20. Nair, V.; Hinton, G.E. "Rectified linear units improve restricted boltzmann machines." In Proceedings of the 27th International Conference on Machine Learning, Haifa, Israel, 807–814(2010).
21. J. Patterson and A. Gibson, "Deep Learning: A Practitioner's Approach." O'Reilly Media,532(2017).
22. F. Chollet, "Keras," https://github.com/fchollet/keras (2015).
23. M. Abadi, A. Agarwal, P. Barham, E. Brevdo, Z. Chen, C. Citro, G. S. Corrado, A. Davis, J. Dean, M. Devin, S. Ghemawat, I. Goodfellow, A. Harp, G. Irving, M. Isard, Y. Jia, R. Jozefowicz, L. Kaiser, M. Kudlur, J. Levenberg, D. Mané, R. Monga, S. Moore, D. Murray, C. Olah, M. Schuster, J. Shlens, B. Steiner, I. Sutskever, K. Talwar, P. Tucker, V. Vanhoucke, V. Vasudevan, F. Viégas, O. Vinyals, P. Warden, M. Wattenberg, M. Wicke, Y. Yu, and X. Zheng, "TensorFlow: large-scale machine learning on heterogeneous systems," software available from http://www.tensorflow.org (2015).
24. D. P. Kingma and J. Ba, "Adam: a method for stochastic optimization," (2014).
25. Sage, D. et al. "Super-resolution fight club: assessment of 2D and 3D single-molecule localization microscopy software." Nature methods 16, 387 (2019).